\documentclass[aps,prl,superscriptaddress,twocolumn]{revtex4}
\usepackage{amssymb}
\usepackage{graphicx}
\usepackage{amsmath}
\usepackage[colorlinks=true,linkcolor=blue,citecolor=blue,urlcolor=blue]{hyperref}

\begin{document}

\title{Maximum Energy Growth Rate in Dilute Quantum Gases }

\author{Ran Qi}
\email{qiran@ruc.edu.cn}
\affiliation{Department of Physics, Renmin University of China, Beijing, 100872, P. R. China}
\author{Zheyu Shi}
\affiliation{Key State Laboratory of Precision Spectroscopy, East China Normal University, Shanghai 200062, China}
\author{Hui Zhai}
\email{hzhai@tsinghua.edu.cn}
\affiliation{Institute for Advanced Study, Tsinghua University, Beijing 100084, China}
\date{\today}

\begin{abstract}

In this letter we study how fast the energy density of a quantum gas can increase in time, when the inter-atomic interaction characterized by the $s$-wave scattering length $a_\text{s}$ is increased from zero with arbitrary time dependence. We show that, at short time, the energy density can at most increase as $\sqrt{t}$, which can be achieved when the time dependence of $a_\text{s}$ is also proportional to $\sqrt{t}$, and especially, a universal maximum energy growth rate can be reached when $a_\text{s}$ varies as $2\sqrt{\hbar t/(\pi m)}$. If $a_\text{s}$ varies faster or slower than $\sqrt{t}$, it is respectively proximate to the quench process and the adiabatic process, and both result in a slower energy growth rate. These results are obtained by analyzing the short time dynamics of the short-range behavior of the many-body wave function characterized by the contact, and are also confirmed by numerical solving an example of interacting bosons with time-dependent Bogoliubov theory. These results can also be verified experimentally in ultracold atomic gases.

\end{abstract}

\maketitle

The ability of tuning interactions between particles is a major advantage of ultracold atomic systems \cite{FR1,FR2}. Especially, by ultilizing magnetic and optical tools, the interaction strength between atoms, usually characterized by the $s$-wave scattering length $a_\text{s}$, can be tuned in a time scale much faster than the many-body relaxation time. This has led to a number of interesting ultracold atomic experiments reported in recent years, such as universal quench dynamics observed by quenching interaction to unitarity \cite{Jin,cambridge}, and coherent excitation of the Higgs model in superfluid Fermi gases and the Bogoliubov quasi-particles in Bose condensate by periodically modulating interactions \cite{Higgs, Cheng1,Cheng2,Cheng3}. These experimental progresses are also accompanied by lots of theoretical interests on studying non-equilibrium dynamics driven by time-dependent interactions \cite{theory1,theory2,theory3,theory4,theory5,theory6,theory7,theory8,theory9,theory10,theory11,theory12,theory13,theory14,theory15,theory16,theory17,theory18,theory19,theory20,theory21,theory22}.

Motivated by these progresses, here we address a fundamental issue whether there is a universal upper limit for the energy increasing rate. To be concrete, suppose that we start with a non-interacting quantum gas with $a_\text{s}=0$ and then change $a_\text{s}$ in time, and suppose that $a_\text{s}(t)$ can be controlled in any function form, the question is whether there is an upper bound for the rate of how fast the total energy can increase as a function of time. In this letter we show that there does exist such a universal rate limit, as far as the initial growth rate is concerned. This result is quite counter-intuitive, because normally the interaction energy increases as the interaction strength increases. Thus, intuitively, one would think that a faster increasing of interaction strength should result in a faster increasing of interaction energy, and consequently, a faster increasing of the total energy. Since we consider that $a_\text{s}$ can be increased as fast as one wants, it seems to indicate that there should not be such a bound.

However, our results show that this intuition is not correct. Before presenting rigorous mathematical statement, we first emphasize that our result is closely tied to a key quantity of ultracold atomic gases called the contact \cite{contact1,contact2,contact3,contact4,contact5,contact6,contact7,bose_contact,contact_pwave}. It is now well known that, for quantum gases with zero-range interactions, one can define contact $\mathcal{C}$ through the short-range behavior of the many-body wave function when any two atoms are brought close to each other, or equivalently, through the high-momentum tail of the momentum distribution. It has been shown that the total energy of a quantum gas is directly related to the contact \cite{contact1,contact2,contact3,contact4,contact5,contact6,contact7,bose_contact,contact_pwave}.

To gain an intuitive understanding of our results, let us first consider two limits. On one limit, the fastest change of the interaction strength is the quench process, during which $a_\text{s}$ instantaneously jumps from zero to any non-zero value. However, it can be shown that the contact does not change and retains zero right after the quench \cite{contact2}, and therefore, the total energy also does not change after the quench. This means that the fastest change of interaction actually does not result in a fast change of the total energy, and in contrast, the interaction energy does not change at all. On the opposite limit, we can consider an adiabatic varying of the interaction strength, during which the interaction energy does vary in time but it varies adiabatically with sufficiently slow rate. The physical pictures in these two limits motivate us to expect a universal maximum growth rate driven by intermediate speed of varying the interaction strength.

\textit{General Expression for the Contact Growth.} Here we consider a uniform Bose gas or spin-1/2 Fermi gas starting from any non-interacting state $|\Psi_0\rangle$ at $t=0$, and then the $s$-wave scattering length $a_\text{s}(t)$ can vary with arbitrary time dependence. Below we use $n$ and $n_\sigma$ to denote the densities of bosons and fermions with spin-$\sigma$ ($\sigma=\uparrow,\downarrow$), respectively, and $\hat{\psi}$ and $\hat{\psi}_\sigma$ to denote boson operator and fermion operator with spin-$\sigma$, respectively. One of the main results of this work states as follows:

In the short-time limit, the dynamics of the contact $\mathcal{C}(t)$ is given by
\begin{equation}
\mathcal{C}(t)=g_2(0)|\eta(t)|^2. \label{Ct}
\end{equation}
Here $g_2({\bf r})$ is defined as $\langle\Psi_0|\hat{\psi}^{\dag}(\frac{\mathbf{r}}{2})\hat{\psi}^{\dag}(-\frac{\mathbf{r}}{2})\hat{\psi}(-\frac{\mathbf{r}}{2})\hat{\psi}(\frac{\mathbf{r}}{2})|\Psi_0\rangle$ for bosons and $\langle\Psi_0|\hat{\psi}^{\dag}_\uparrow(\frac{\mathbf{r}}{2})\hat{\psi}^{\dag}_\downarrow(-\frac{\mathbf{r}}{2})\hat{\psi}_\downarrow(-\frac{\mathbf{r}}{2})\hat{\psi}_\uparrow(\frac{\mathbf{r}}{2})|\Psi_0\rangle$ for fermions, and $g_2(0)$ means $g_2({\bf r})$ evaluated at ${\bf r}=0$. Especially, if $|\Psi_0\rangle$ is the non-interacting ground state, then $g_2(\mathbf{r})=n^2$ or $n_{\uparrow}n_{\downarrow}$ for the Bose or the spin-$1/2$ Fermi gas. The key result is that the function $\eta(t)$ obeys the following integral equation
\begin{eqnarray}
\left[\hat{L}+\frac{1}{4\pi a_\text{s}(t)}\right]\eta(t)=-1,\label{freeEta}
\end{eqnarray}
where $\hat{L}$ is a linear operator acting on $\eta(t)$ as
\begin{align}
\hat{L}\eta(t)=&\left(\frac{m}{\hbar}\right)^{\frac{1}{2}}\frac{1}{8\pi^{3/2}\sqrt{i}}\times \nonumber\\
&\lim_{\epsilon\rightarrow0^+}\left[\int_0^{t-\epsilon} \frac{\eta(\tau)}{(t-\tau)^{\frac{3}{2}}}d\tau-\frac{2\eta(t)}{\sqrt{\epsilon}}\right].
\end{align}

This result is motivated by solving the two-body problem, which satisfies the following Schr$\ddot{\text{o}}$dinger equation in the relative coordinate ${\bf r}$ frame as
\begin{eqnarray}
i\hbar\frac{\partial \psi}{\partial t}=-\frac{\hbar^2\nabla^2 \psi}{m}+\frac{4\pi\hbar^2 a_\text{s}(t)}{m}\delta(\mathbf{r})\frac{\partial}{\partial r}r\psi.
\end{eqnarray}
Starting from an initial state $\psi({\bf r})=1/\sqrt{V}$ ($V$ is the total volume of the system), the time evolution of the wave function  always obeys the following asymptotic form in the short-range $r\rightarrow 0$ limit, that is
\begin{eqnarray}
\psi(\mathbf{r},t)=\frac{\eta(t)}{4\pi\sqrt{V}}\left[\frac{1}{r}-\frac{1}{a_\text{s}(t)}\right]+O(r),
\end{eqnarray}
and it can be shown that $\eta(t)$ satisfies Eq.~\ref{freeEta} \cite{supple}. Generalizing this result from the two-body problem to the many-body problem utilizes the short-time expansion and is quite straightforward, which yields Eq.~\ref{Ct} \cite{supple}. Here we note that for the two-body problem, $\eta(t)$ satisfies Eq.~\ref{freeEta} for all time scales, but for the many-body problem, the result is only valid for the short-time scale. Here short-time is defined as the time scale much shorter than the typical many-body time scale $t_n=\hbar/E_\text{n}$, where $E_\text{n}=\hbar^2k_\text{n}^{2}/(2m)$ and $k_n=(6\pi^2 n)^{1/3}$ (with $n$ replaced by $n_\sigma$ for fermions). In other word, in such short time scale, the short-range behavior of many-body wave function is still dominated by the two-body physics.

\begin{figure}[t]
    \centering
    \includegraphics[width=0.48\textwidth]{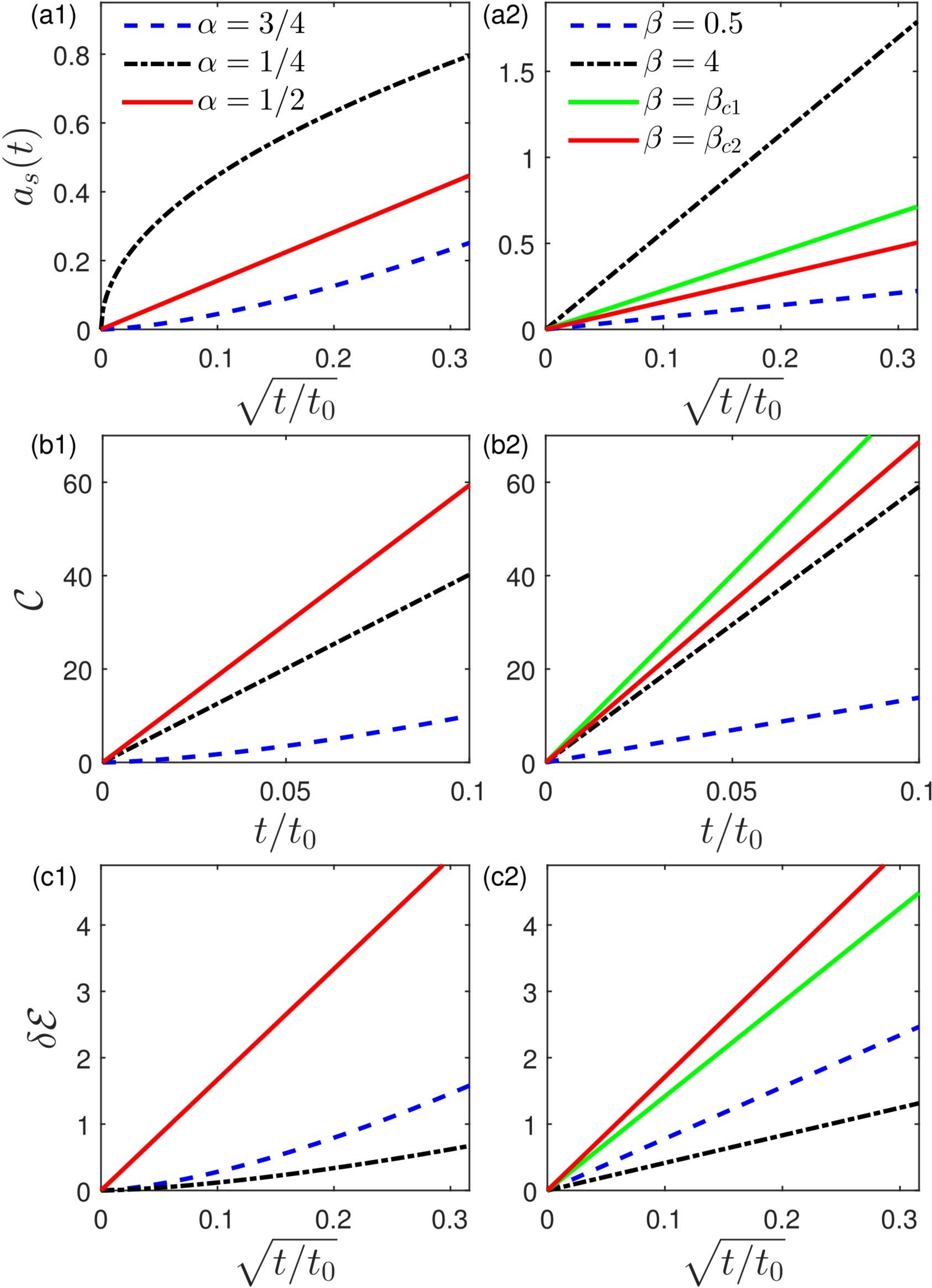}
    \caption{(a1-a2) The time-dependence of the scattering length $a_\text{s}(t)$ (in unit of $l_0$) with different power-law functions of Eq.~\ref{a_s_power_law}. (a1) $\alpha=\frac{1}{4},~\frac{1}{2},~\frac{3}{4}$, and $\beta$ is fixed at $\beta=1$. (a2) $\alpha$ is fixed at $\alpha=\frac{1}{2}$ and $\beta=\frac{1}{2},~1.128,~1.596$ and $4$. (b1-b2) The short time behavior of the contact $\mathcal{C}$ (in unit of $g_2(0)l_0^2$) with $a_\text{s}(t)$ plotted in (a1) and (a2), respectively. (c1-c2) The time-dependence of the energy density change $\delta\mathcal{E}$ (in unit of $g_2(0)l_0\hbar^2/m$) with $a_\text{s}(t)$ plotted in (a1) and (a2), respectively. }
     \label{growth}
\end{figure}

\textit{Contact Growth Rate.} Here, without loss of generality, we consider that $a_\text{s}(t)$ grows from zero to a positive value in a power-law function as
\begin{equation}
a_\text{s}(t)=\sqrt{2}\beta l_0\left(\frac{t}{t_0}\right)^\alpha, \label{a_s_power_law}
\end{equation}
where $l_0$ is an arbitrary length unit and $t_0$ is the time units, and $l_0$ and $t_0$ are both related to the same energy unit as $\hbar/t_0=\hbar^2/(2ml^2_0)$. $\alpha,\beta$ are two constants describing the power and the coefficient, respectively and a factor $\sqrt{2}$ is introduced just for the later convenience. The operator $\hat{L}$ has an important property that
\begin{eqnarray}
\hat{L}t^{\alpha}=\left(\frac{m}{\hbar}\right)^{\frac{1}{2}}B(\alpha)t^{\alpha-\frac{1}{2}},\label{Leta}
\end{eqnarray}
where $B(\alpha)$ is a constant given by $B(\alpha)=i^{3/2}\Gamma(\alpha+1)/(4\pi\Gamma(\alpha+1/2))$. That is to say, suppose $\eta(t)$ is a power-law function in $t$, when $\hat{L}$ acts on $\eta(t)$, it lowers the power of $\eta(t)$ by $1/2$. This property plays a crucial role in the following analysis because it means whether $\alpha$ in Eq. \ref{a_s_power_law} is greater or smaller than $1/2$ makes significant difference.

Case I: $\alpha>1/2$. In this case, the $1/(4\pi a_\text{s})$ term dominates Eq.~\ref{freeEta}, and thus, to the leading order of $t$, $\eta(t)$ and $\mathcal{C}(t)$ are given by
\begin{equation}
\eta(t)=-4\pi a_\text{s}(t);  \   \   \mathcal{C}(t)=16\pi^2 a^2_\text{s}(t) g_2(0) .  \label{case1}
\end{equation}
This is consistent with the adiabatic regime where the physical quantities only depend on the instantaneous scattering length at time $t$.

Case II: $\alpha<1/2$. In this case, the $\hat{L}$ term dominates Eq.~\ref{freeEta}, and thus, to the leading order of $t$, $\eta(t)$ and $\mathcal{C}(t)$ are given by
\begin{equation}
\eta(t)=-\left(\frac{\hbar}{m}\right)^{\frac{1}{2}}\frac{1}{B(1/2)}\sqrt{t};  \   \ \mathcal{C}(t)=\frac{\hbar}{m}\frac{g_2(0)}{B(1/2)^2}t, \label{case2}
\end{equation}
where $B(1/2)=-1/(8\sqrt{i\pi})$. Surprisingly, in this case this result shows that the growth of contact at the short-time is independent of parameters $l_0$, $t_0$, $\alpha$ and $\beta$ in Eq. \ref{a_s_power_law}. That is to say, it is independent of how fast $a_\text{s}$ varies in time. Even if $l_0$ or $\beta$ is infinitely large, or $\alpha$ is infinitesimally small, and then $a_\text{s}(t)$ initially grows infinitely fast, the contact always grows linearly in time with a constant rate. This means that as long as $\alpha<1/2$, the short-range physics at the short time is the same as a quench process where the scattering length instantaneously jumps to unitarity.

Case III: $\alpha=1/2$. In this case, Eq. \ref{a_s_power_law} becomes
\begin{equation}
a_\text{s}(t)=\beta\sqrt{\frac{\hbar t}{m}},
\end{equation}
By dimension analysis, it is easy to see that $l_0$ and $t_0$ cancel each other and only the coefficient $\beta$ enters the expression.
 In this case, both $\hat{L}$ term and the $1/(4\pi a_\text{s})$ term are equally important. Also to the leading order of $t$, we obtain
\begin{equation}
\eta(t)=-A(\beta)\sqrt{t};  \   \   \mathcal{C}(t)=|A(\beta)|^2 g_2(0)t, \label{case3}
\end{equation}
where $A(\beta)$ is also a constant given by
\begin{equation}
A(\beta)=\left(\frac{\hbar}{m}\right)^{\frac{1}{2}}\frac{1}{B\left(\frac{1}{2}\right)+\frac{1}{4\pi \beta}}.
\end{equation}
As one can see from here, this is a critical case. In the Case III, by taking $\beta\rightarrow \infty$, Eq.~\ref{case3} recovers Eq.~\ref{case2}, consistent with the quench limit, and by taking $\beta\rightarrow 0$, Eq.~\ref{case3} recovers Eq.~\ref{case1}, consistent with the adiabatic limit.

Here an important point is that $|A(\beta)|^2$ is \textit{not} a monotonic function in $\beta$. For a given initial state, $g_2(0)$ is fixed, and we can then define the initial growth rate for contact as $v_\mathcal{C}=\lim_{t\rightarrow 0}d\mathcal{C}(t)/dt$. As one can see from Eq.~\ref{case1}, $v_\mathcal{C}=0$ for case I. And for both the Case II and the Case III, $v_\text{C}$ is a constant, given by $|A(\beta)|^2g_2(0)$ for the Case III and $|A(\beta=\infty)|^2g_2(0)$ for the Case II. It turns out that $|A(\beta)|^2$ reaches its maximum at $\beta_{\text{c1}}=2\sqrt{2/\pi}\approx 1.596$, at which $v^\text{max}_\mathcal{C}=(\hbar/m)128\pi g_2(0)$.

\begin{figure}[t]
    \centering
    \includegraphics[width=0.45\textwidth]{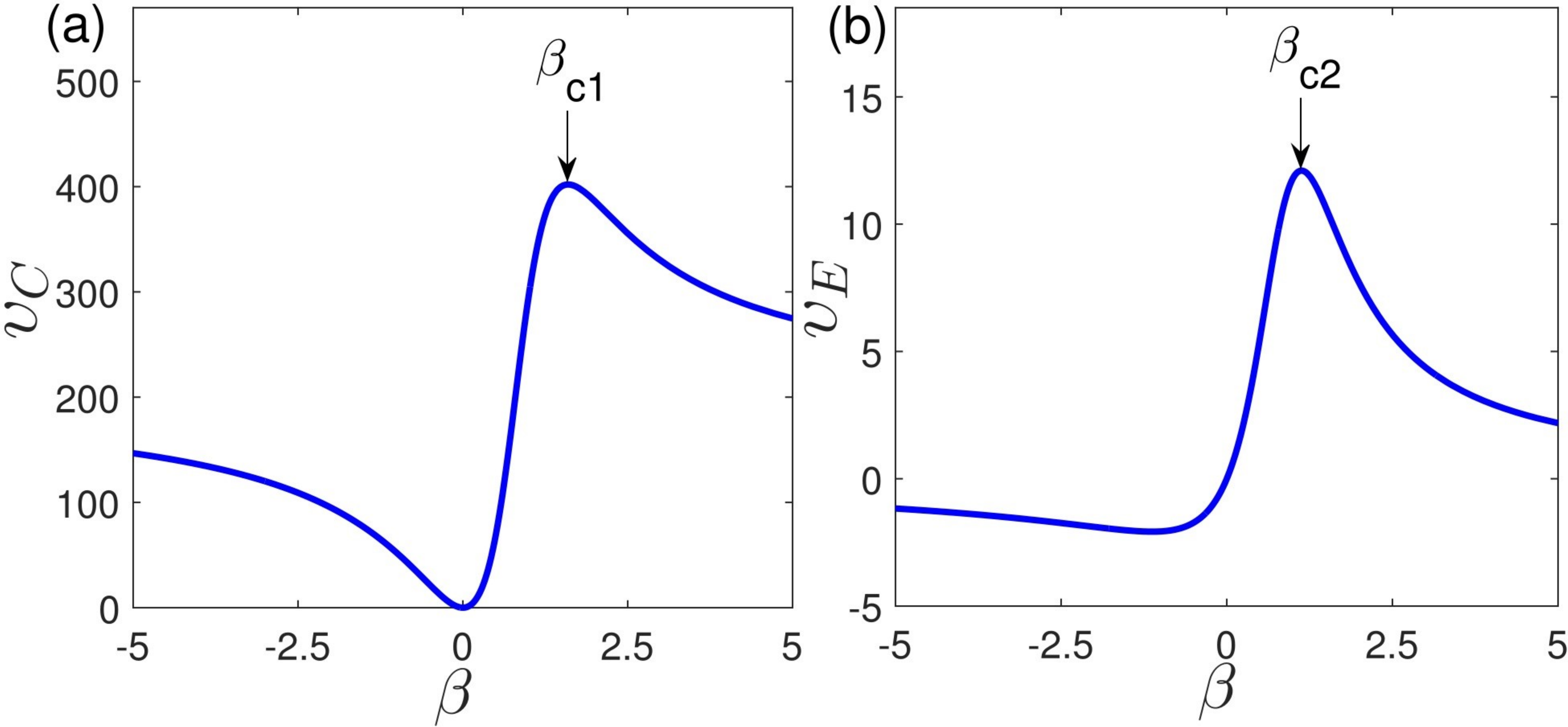}
    \caption{Initial growth rate for contact (a) and for energy (b) as a function of $\beta$ for $a_\text{s}(t)= \beta \sqrt{\hbar t/m}$. Arrows mark $\beta_{\text{c1}}$ and $\beta_{\text{c2}}$ where the maximum contact growth rate and the maximum energy growth rate are reached. $\upsilon_C$ and $\upsilon_E$ are plotted in units of $g_2(0)\hbar/m$ and $g_2(0)\sqrt{\hbar^3/m}$ respectively. }
     \label{rate}
\end{figure}

\textit{Energy Growth Rate.} The total energy density of a uniform zero-range interacting quantum gas can be measured through its momentum distribution $n_{{\bf k}}$. For example, for spin-$1/2$ fermions, it is given by
\begin{equation}
\mathcal{E}=\int\frac{d^3k}{(2\pi)^3}\epsilon_{{\bf k}}\left(n_{{\bf k}}-\frac{2\mathcal{C}}{k^4}\right)+\frac{\mathcal{C}}{4\pi m a_s},\label{energy_expression}
\end{equation}
where $n_{{\bf k}}=n_{{\bf k}\uparrow}+n_{{\bf k}\downarrow}$, $\epsilon_{{\bf k}}=\hbar^2{\bf k}^2/(2m)$, and the contact $\mathcal{C}$ is related to $n_{{\bf k}\sigma}$ through $\mathcal{C}\equiv\lim_{k\rightarrow \infty}k^4n_{{\bf k}\sigma}$ \cite{contact1}. The same expression, replacing all $\mathcal{C}$ by $\mathcal{C}/2$, also holds for the spinless Bose gas as long as the three-body contact can be ignored \cite{bose_contact}.

On the other hand, there is a direct relation between the time evolution of the energy and the contact. For spin-$1/2$ fermions it is given as
\begin{eqnarray}
\frac{d}{dt}\mathcal{E}(t)=\frac{ \hbar^2 \mathcal{C}(t)}{4\pi m a_\text{s}^2(t)}\frac{da_\text{s}}{dt}\label{Et}.
\end{eqnarray}
For spinless bosons, an extra $1/2$ factor should also be added in the r.h.s. of Eq. \ref{Et}.
Therefore, based on the contact growth discussed above, we can determine the energy growth.

Case I: $\alpha>1/2$. With the help of Eq. \ref{case1}, one can obtain that
\begin{equation}
\delta \mathcal{E}(t)=\frac{4\pi\hbar^2 a_\text{s}(t)}{m}g_2(0).
\end{equation}
where $\delta\mathcal{E}(t)=\mathcal{E}(t)-\mathcal{E}(t=0)$. This result again shows that the physics in this regime is consistent with adiabatic regime where the energy is determined by the instantaneous scattering length. Since $\alpha>1/2$, the energy increases slower than $\sqrt{t}$ at the short time.

Case II: $\alpha<1/2$. In this regime, Eq. \ref{case2} gives rise to
\begin{equation}
\delta \mathcal{E}(t)=\frac{16\sqrt{2}\alpha}{\beta(1-\alpha)}\frac{\hbar^2}{m} g_2(0)l_0\left(\frac{t}{t_0}\right)^{1-\alpha}.
\end{equation}
Since $\alpha<1/2$, the energy also increases slower than $\sqrt{t}$ at the short time. When taking the $\alpha\rightarrow 0$ limit, or $\beta\rightarrow\infty$ or $l_0\rightarrow\infty$ limit, $\delta\mathcal{E}(t)\rightarrow 0$, and it is consistent with the fact there is no energy change for the quench process as discussed above.

Case III: $\alpha=1/2$. In this regime, Eq. \ref{case3} yields
\begin{equation}
\delta \mathcal{E}(t)=\sqrt{\frac{\hbar^3}{m}}\frac{|A(\beta)|^2}{4\pi\beta}g_2(0)\sqrt{t}.
\end{equation}
It is in this case that the energy growth at the short time is the fastest. Now we can define an energy growth rate $v_E=\lim_{t\rightarrow 0}d\mathcal{E}(t)/d\sqrt{t}$. For case I and II, this rate is zero. In case II, $v_E$ is given by $\sqrt{\hbar^3/m}|A(\beta)|^2g_2(0)/(4\pi\beta)$, which reaches its maximum at $\beta_{\text{c2}}=2/\sqrt{\pi}\approx1.128$ with $v^\text{max}_E=4(2+\sqrt{2})\sqrt{\pi}g_2(0)\sqrt{\hbar^3/m}\approx 24.2g_2(0)\sqrt{\hbar^3/m} $. Note that this value of $v^\text{max}_E$ applies for the spin-$1/2$ Fermi gas, and for the spinless Bose gas an extra 1/2 factor should be added.

This maximum energy growth rate is the main result of this work. We note that, although this result is obtained by assuming power-law function of $a_\text{s}(t)$ and by considering positive $a_\text{s}(t)$, it can be extended to other function forms, such as including the logarithmic function corrections, and including the situations where $a_\text{s}$ varies to negative values. The results discussed above are summarized in Fig. \ref{growth} and Fig. \ref{rate}. Fig. \ref{growth}(a1) and (a2) show different power-law function of $a_\text{s}(t)$ given by Eq.~\ref{a_s_power_law}, either with different power $\alpha$, or with different coefficient $\beta$ and fixed $\alpha=1/2$. Fig. \ref{growth}(b1) and (b2) show the corresponding contact growth, and Fig. \ref{growth}(c1) and (c2) show the corresponding energy growth, using spinless bosons as an example. It clearly shows that a faster increasing of $a_\text{s}$ does not necessarily lead to a faster increasing of the contact and the energy density. One can see that for different powers, $\alpha=1/2$ gives the fastest contact growth and energy growth at the short time. And for $\alpha$ fixed at $1/2$, $\beta=\beta_{\text{c1}}$ yields the fastest contact growth and $\beta=\beta_{\text{c2}}$ yields the fastest energy growth, as also shown in Fig. \ref{rate}.

\begin{figure}[t]
    \centering
    \includegraphics[width=0.47\textwidth]{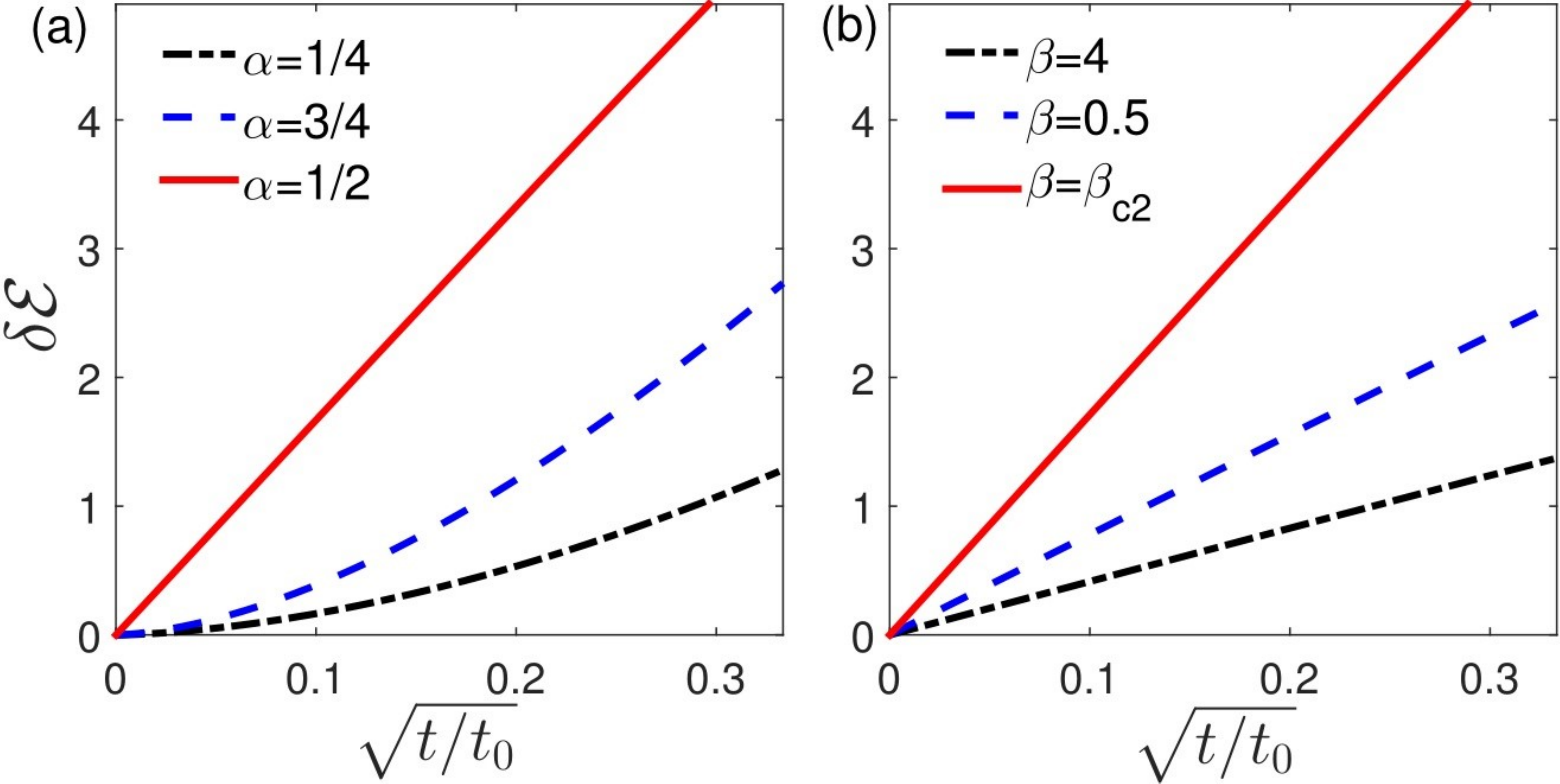}
    \caption{Dynamics of the total energy density of Bose gas for $a_{\text{s}}(t)$ with different power-law functions of Eq.~\ref{a_s_power_law}. (a) $\beta=1$ and $\alpha=\frac{1}{4},~\frac{1}{2},~\frac{3}{4}$. (b) $\alpha=\frac{1}{2}$ and $\beta=4,~0.5,~1.128$. $\delta\mathcal{E}$ is plotted in unit of $n^2l_0\hbar^2/m$ and we have set $t_0=t_n$ and thus $l_0=1/k_n$ in the numerical calculation.}
     \label{boseenergy}
\end{figure}

\textit{Example.} The analysis above is based on the short time expansion. Here, we consider a concrete example of spinless bosons, which can be described by the following time-dependent Hamiltonian
\begin{eqnarray}
\hat{H}(t)=\sum_{\mathbf{k}}\epsilon_{\mathbf{k}}\hat{b}_{\mathbf{k}}^{\dag}\hat{b}_{\mathbf{k}}+
\frac{U(t)}{2 V}\sum_{\mathbf{k},\mathbf{k}',\mathbf{q}}\hat{b}_{\mathbf{k}}^{\dag}\hat{b}_{\mathbf{q}-\mathbf{k}}^{\dag}\hat{b}_{\mathbf{q}-\mathbf{k}'}\hat{b}_{\mathbf{k}'},
\end{eqnarray}
where $\hat{b}_{{\bf k}}$ are boson creation operators with momentum ${\bf k}$. $U(t)$ is related to $a_\text{s}(t)$ through the renormalization  relation
\begin{equation}
\frac{1}{U(t)}=\frac{m}{4\pi\hbar^2 a_\text{s}(t)}-\frac{1}{V}\sum_{\mathbf{k}}\frac{1}{2\epsilon_{\mathbf{k}}}.
\end{equation}
We solve this Hamiltonian by adopting the Bogoliubov-type variational ansatz as
\begin{eqnarray}
|\Psi(t)\rangle=\mathcal{N}(t)\exp\left[g_0(t) \hat{b}^\dag_0+\sum_{\mathbf{k}\neq 0}g_{\mathbf{k}}(t)\hat{b}_{\mathbf{k}}^{\dag}\hat{b}_{-\mathbf{k}}^{\dag} \right]|0\rangle,
\end{eqnarray}
where $\mathcal{N}(t)$ is a normalization factor, $|0\rangle$ is vacuum of particles, and $g_0$ and $g_{{\bf k}}$ are all variational parameters. This approach is not restricted to the short time and has been successfully used in the previous studies of degenerate Bose gas quenched to unitarity \cite{theory2,theory5,theory18}. The evolution of variational parameters $g_0(t)$ and $g_{\mathbf{k}}(t)$ can be obtained from the Euler-Lagrange equation for the
Lagrangian $\mathcal{L}=\frac{1}{2}[\langle\Psi(t)|\dot{\Psi}(t)\rangle-\langle\dot{\Psi}(t)|\Psi(t)\rangle]-\langle\Psi(t)|\hat{H}(t)|\Psi(t)\rangle$, which yields a set of differential equations for $g_0$ and $g_{{\bf k}}$. Since we start with a non-interacting Bose condensate, we take $g_0=1$ and $g_{{\bf k}}=0$ at $t=0$ as the initial conditions for these differential equations. We can obtain the variational wave function by solving these equations, and subsequently, we can determine the total energy density with Eq. \ref{energy_expression}. The results for the total energy density are shown in Fig. \ref{boseenergy} for different powers and different coefficients. One can see that the short time behaviors agree very well with that given in Fig. 1(c1) and (c2).

\textit{Summary.} In summary, we have studied the energy growth rate of degenerate quantum gas driven by increasing the $s$-wave scattering length $a_\text{s}$ from zero, by both analyzing the short time behavior on general situations and numerically solving a concrete example of interacting bosons. Two main results are summarized as follows: (i) For energy increasing as $t^\alpha$ at the short time, $\alpha$ cannot be smaller than $1/2$ and $\alpha=1/2$ is achieved when $a_\text{s}(t)$ varies as $\propto\sqrt{t}$. (ii) For energy increasing as $\sqrt{t}$ at the short time, the fastest energy increasing is achieved when $a_\text{s}(t)=2\sqrt{\hbar t/(\pi m)}$, with a maximum energy growth given by $4(2+\sqrt{2})\sqrt{\pi\hbar^3t/m}g_2(0)$ for the spin-$1/2$ Fermi gas and half of that for the spinless Bose gas. This prediction can be directly verified in cold atom experiments. We emphasize that this maximum energy growth rate is universal, that is, it is independent of any length or energy scale. This is because when $a_\text{s}$ varies as $\sqrt{t}$, the entire many-body Schr\"odinger equation is invariant under a space-time scaling transformation $t\rightarrow \lambda^2 t$ and $r\rightarrow \lambda r$.  Similar examples of such scale invariant many-body dynamics have been studied in \cite{Efimovian,Efimovian2,Efimovian3}. Hence, this result ties together the fastest energy growth with the scaling symmetry, and this is reminiscent of an equilibrium analogy, where the interaction effect is the strongest at unitarity where the system is also scale invariant.

\textit{Acknowledgment.}
We thank Peng Zhang for helpful discussions. The project was supported by NSFC under Grant No. 12022405 (RQ), No. 11774426 (RQ) and No.  11734010 (HZ and RQ), Beijing Outstanding Young Scholar Program (HZ), the National Key R and D Program of China under Grant No. 2018YFA0306501(RQ), the Research Funds of Renmin University of China under Grant No. 19XNLG12 (RQ), the Beijing Natural Science Foundation under Grant No. Z180013 (RQ), and Program of Shanghai Sailing Program Grant No. 20YF1411600 (ZYS).

\begin{widetext}
\newpage
\section*{Supplementary material: \\
Maximum Energy Growth Rate in Dilute Quantum Gases}

In this supplementary material, we provide the details of proof for Eq. (1)-(3) in the main text.

\section{solution of two-body problem}
In this section, we solve the time-dependent two-body problem and establish Eq. (2) and (3). We consider the following time-dependent Schr\"odinger equation in the relative coordinate ${\bf r}$ frame
\begin{eqnarray}
i\hbar\frac{\partial }{\partial t}\psi(\mathbf{r},t)=-\frac{\hbar^2}{m}\nabla^2 \psi(\mathbf{r},t)+\frac{4\pi\hbar^2 a_\text{s}(t)}{m}\delta(\mathbf{r})\frac{\partial}{\partial r}r\psi(\mathbf{r},t),\label{2bschr}
\end{eqnarray}
which is Eq. (4) in the main text and we choose the initial state $\psi(\mathbf{r})=1/\sqrt{V}$. We first define an auxiliary function $\eta(t)$ as
\begin{eqnarray}
\eta(t)=-\sqrt{V}4\pi a_s(t)\lim_{r\rightarrow 0}\frac{\partial}{\partial r}[r\psi(\mathbf{r},t)].\label{etadef}
\end{eqnarray}
Then Eq. (\ref{2bschr}) can be rewritten as
\begin{eqnarray}
i\hbar\frac{\partial }{\partial t}\psi(\mathbf{r},t)=-\frac{\hbar^2}{m}\nabla^2 \psi(\mathbf{r},t)-\frac{1}{\sqrt{V}}\frac{\hbar^2}{m}\delta(\mathbf{r})\eta(t).\label{Schr}
\end{eqnarray}
Now Eq. (\ref{Schr}) can be solved with the standard Green's function approach and the solution is given as
\begin{eqnarray}
\psi(\mathbf{r},t)&=&\psi_0(\mathbf{r},t)+ \frac{i}{\sqrt{V}}\frac{\hbar}{m}\int_0^t G_0^{\text{rel}}(\mathbf{r},t-\tau)\eta(\tau)\tau\label{twobodysolution}
\end{eqnarray}
where $G_0^{\text{rel}}(\mathbf{r},t-\tau)$ is the noninteracting Green's function in the relative coordinate frame given by
\begin{eqnarray}
G_0^{\text{rel}}(\mathbf{r},t-\tau)=\left[\frac{m}{i4\pi\hbar (t-\tau)}\right]^{3/2}\exp\left[i\frac{mr^2}{4\hbar(t-\tau)}\right],
\end{eqnarray}
and $\psi_0(\mathbf{r},t)$ satisfies the non-interacting Schr\"odinger equation
\begin{eqnarray}
i\hbar\frac{\partial}{\partial t}\psi_0(\mathbf{r},t)=-\frac{\hbar^2}{m}\nabla^2 \psi_0(\mathbf{r},t),~~~~~~\psi_0(\mathbf{r},t=0)=\psi(\mathbf{r})
\end{eqnarray}
where $\psi(\mathbf{r})$ is the initial wave function of $\psi(\mathbf{r},t)$.

For $\psi(\mathbf{r})=1/\sqrt{V}$, we have $\psi_0(\mathbf{r},t)=1/\sqrt{V}$ and, based on Eq. (\ref{twobodysolution}), we obtain the following asymptotic expansion at $r\rightarrow0$
\newpage
\begin{eqnarray}
\psi(\mathbf{r},t)&=&\frac{1}{\sqrt{V}}\left\{1+\frac{\eta(t)}{4\pi r}+ \left(\frac{m}{\hbar}\right)^{\frac{1}{2}}\frac{1}{8\pi^{3/2}\sqrt{i}} \lim_{r\rightarrow0^+}\left[\int_0^{t} \frac{\eta(\tau)\exp\left[i\frac{mr^2}{4\hbar(t-\tau)}\right]}{(t-\tau)^{\frac{3}{2}}}d\tau-\int_{-\infty}^{t} \frac{\eta(t)\exp\left[i\frac{mr^2}{4\hbar(t-\tau)}\right]}{(t-\tau)^{\frac{3}{2}}}d\tau\right]+\text{O}(r)\right\},\nonumber\\
&=&\frac{1}{\sqrt{V}}\left\{1+\frac{\eta(t)}{4\pi r}+ \left(\frac{m}{\hbar}\right)^{\frac{1}{2}}\frac{1}{8\pi^{3/2}\sqrt{i}} \lim_{\epsilon\rightarrow0^+}\left[\int_0^{t-\epsilon} \frac{\eta(\tau)}{(t-\tau)^{\frac{3}{2}}}d\tau-\int_{-\infty}^{t-\epsilon} \frac{\eta(t)}{(t-\tau)^{\frac{3}{2}}}d\tau\right]+\text{O}(r)\right\},\nonumber\\
&=&\frac{1}{\sqrt{V}}\left\{1+\frac{\eta(t)}{4\pi r}+ \left(\frac{m}{\hbar}\right)^{\frac{1}{2}}\frac{1}{8\pi^{3/2}\sqrt{i}} \lim_{\epsilon\rightarrow0^+}\left[\int_0^{t-\epsilon} \frac{\eta(\tau)}{(t-\tau)^{\frac{3}{2}}}d\tau-\frac{2\eta(t)}{\sqrt{\epsilon}}\right]+\text{O}(r)\right\},\nonumber\\
&=&\frac{1}{\sqrt{V}}\left[1+\frac{\eta(t)}{4\pi r}+ \hat{L}\eta(t)+\text{O}(r)\right],\label{BP2}
\end{eqnarray}
where $\hat{L}$ is defined in Eq. (3) in the main text. Then substituting Eq. (\ref{BP2}) into the r.h.s. of Eq. (\ref{etadef}), we immediately  obtain
\begin{eqnarray}
\left[\hat{L}+\frac{1}{4\pi a_\text{s}(t)}\right]\eta(t)=-1,\label{freeEta}
\end{eqnarray}
which is exactly the Eq. (2) in the main text.

\section{Generalization to many-body wave function}\label{general solution}

In this section, we generalize the two-body solution to many-body wave functions and establish Eq. (1) in the main text. To simplify notations, we set $\hbar=m=1$ in this section. Since we are only interested in the short time evolution of the short distance behavior of the many-body wave function, it is convenient to divide the entire $3N$ dimensional space into two regions: $\mathcal{D}_{\epsilon}$ and $\mathcal{I}_{\epsilon}$. $\mathcal{D}_{\epsilon}$ denotes the region in which the distance between any two bosons is larger than $\epsilon$, and $\mathcal{I}_{\epsilon}$ is the complementary space to $\mathcal{D}_{\epsilon}$. In the short time limit $t\ll t_n$, it is possible to choose an intermediate length scale $\epsilon$ such that $\sqrt{t}\ll\epsilon\ll 1/k_n$ ($t_n$ and $k_n$ was already defined in the main text). Such dividing of configuration space will be very useful in the following proof of this section.

We consider the following many-body Schr$\ddot{\text{o}}$inger equation for N-interacting bosons with a time-dependent s-wave scattering length $a_\text{s}(t)$
\begin{eqnarray}
i\frac{\partial}{\partial t}\psi(\mathbf{r}_1,\mathbf{r}_2,\cdots\mathbf{r}_N;t)=-\frac{1}{2}\sum_{i=1}^N\nabla^2_i\psi(\mathbf{r}_1,\mathbf{r}_2,\cdots\mathbf{r}_N;t)+4\pi a_\text{s}(t)\sum_{i<j}\delta(\mathbf{r}_{ij})\frac{\partial }{\partial r_{ij}}r_{ij}\psi(\mathbf{r}_1,\mathbf{r}_2,\cdots\mathbf{r}_N;t)\label{schroMB}
\end{eqnarray}
where $\mathbf{r}_{ij}=\mathbf{r}_i-\mathbf{r}_j$ and $\psi(\mathbf{r}_1,\mathbf{r}_2,\cdots\mathbf{r}_N;t)$ is the normalized N-body wave function satisfying
\begin{equation}
\int d^3\mathbf{r}_1\,\cdots d^3\mathbf{r}_N|\psi(\mathbf{r}_1,\mathbf{r}_2,\cdots\mathbf{r}_N;t)|^2=1.
\end{equation}

We first define an auxiliary function
\begin{eqnarray}
\Phi(\mathbf{R};\mathbb{Z};t)=-4\pi a_\text{s}(t)\lim_{\mathbf{r}\rightarrow0}\frac{\partial }{\partial r}r\psi(\mathbf{R}+\mathbf{r}/2,\mathbf{R}-\mathbf{r}/2,\mathbf{z}_1,\cdots\mathbf{z}_{N-2};t)\label{Phi}
\end{eqnarray}
where $\mathbb{Z}=\mathbf{z}_1,\cdots\mathbf{z}_{N-2}$. Then based on Eq. (\ref{schroMB}) and (\ref{Phi}), one can show that the N-body wave function satisfies the following asymptotic boundary condition at $r_{ij}\rightarrow0$ with any $i<j$:
\begin{eqnarray}
\psi(\mathbb{R};t)=\frac{1}{4\pi}\Phi(\mathbf{R}_{ij};\mathbb{R}_{\overline{ij}};t)\left(\frac{1}{r_{ij}}-\frac{1}{a_\text{s}(t)} \right)+O(r_{ij}),
\end{eqnarray}
where $\mathbf{R}_{ij}=(\mathbf{r}_i+\mathbf{r}_j)/2$ and, to simplify notations, we have defined $\mathbb{R}=\mathbf{r}_1,\mathbf{r}_2,\cdots\mathbf{r}_N$ and $\mathbb{R}_{\overline{ij}}=\mathbf{r}_1,\cdots,\mathbf{r}_{i-1},\mathbf{r}_{i+1},\cdots,\mathbf{r}_{j-1},\mathbf{r}_{j+1},\cdots\mathbf{r}_N$ denotes the coordinates excluding ${\bf r}_i$ and ${\bf r}_j$.

The function $\Phi(\mathbf{R};\mathbb{Z};t)$ is directly related to the contact as
\begin{eqnarray}
\mathcal{C}(t)&=&\frac{1}{V}16\pi^2N(N-1)\int d^3\mathbf{R}d^{3(N-2)}\mathbb{Z}\left|\frac{1}{4\pi}\Phi(\mathbf{R};\mathbb{Z};t)\right|^2,\label{contact}
\end{eqnarray}
where $V$ is the total volume of system and $d^{3(N-2)}\mathbb{Z}=d^3\mathbf{z}_1\,\cdots d^3\mathbf{z}_{N-2}$.

Now Eq. (\ref{schroMB}) can be rewritten as
\begin{eqnarray}
i\frac{\partial}{\partial t}\psi(\mathbb{R};t)=-\frac{1}{2}\sum_{i=1}^N\nabla^2_i\psi(\mathbb{R};t)
-\sum_{i<j}\delta(\mathbf{r}_{ij})\Phi\left(\frac{\mathbf{r}_i+\mathbf{r}_j}{2};\mathbb{R}_{\overline{ij}};t\right).
\end{eqnarray}

This equation can still be solved by standard Green's function approach and we obtain the following form of solution
\begin{eqnarray}
&&\psi(\mathbb{R};t)=\psi_0(\mathbb{R};t)+\sum_{i<j}\psi_{\text{int}}(\mathbf{r}_i,\mathbf{r}_j;\mathbb{R}_{\overline{ij}};t),\label{psi}\\
&&\psi_{\text{int}}(\mathbf{a},\mathbf{b};\mathbb{Z};t)=i\int_0^td\tau G_0^{\text{rel}}(\mathbf{a}\!-\!\mathbf{b},t\!-\!\tau)\Phi_0^{t-\tau}\left(\frac{\mathbf{a}\!+\!\mathbf{b}}{2},\mathbb{Z};\tau\right),\\
&&\Phi_0^{t-\tau}(\mathbf{R};\mathbb{Z};\tau)=\int d^3\mathbf{R}'d^{3(N-2)}\mathbb{Z}'G_0^{\text{cm}}\left(\mathbf{R}-\mathbf{R}',t\!-\!\tau\right)\frac{\exp\left[i\frac{(\mathbb{Z}-\mathbb{Z}')^2}{2(t-\tau)}\right]}{[2\pi i(t-\tau)]^{3(N-2)/2}}\Phi(\mathbf{R}';\mathbb{Z}';\tau),
\end{eqnarray}
where we defined the following two Green's functions
\begin{equation}
G_0^{\text{cm}}(\mathbf{r},t)=\frac{1}{(\pi it)^{3/2}}\exp\left(i\frac{r^2}{t}\right),~~~~~~
G_0^{\text{rel}}(\mathbf{r},t)=\frac{1}{(4\pi it)^{3/2}}\exp\left(i\frac{r^2}{4t}\right).
\end{equation}
The function $\Phi_0^{t-\tau}(\mathbf{R};\mathbb{Z};\tau)$ can be seen as the free expansion with time $t-\tau$ from an `initial wave function' $\Phi(\mathbf{R};\mathbb{Z};\tau)$.

The non-interacting wave function $\psi_0(\mathbb{R};t)$ satisfies
\begin{eqnarray}
i\frac{\partial}{\partial t}\psi_0(\mathbb{R};t)=-\frac{1}{2}\sum_{i=1}^N\nabla^2_i\psi_0(\mathbb{R};t),~~~~~~~~\psi_0(\mathbb{R};t=0)=\psi(\mathbb{R};t=0).
\end{eqnarray}

It is easy to show that $\psi_{\text{int}}(\mathbf{a},\mathbf{b};\mathbb{Z};t)$ is only singular at $\mathbf{a}-\mathbf{b}\rightarrow0$, and we obtain the following asymptotic behavior as $r_{ij}\equiv|{\bf r}_{ij}|\rightarrow0$
\begin{eqnarray}
\psi_{\text{int}}(\mathbf{r}_i,\mathbf{r}_j;\mathbb{R}_{\overline{ij}};t)=\frac{\Phi(\mathbf{R}_{ij},\mathbb{R}_{\overline{ij}};t)}{4\pi r_{ij}}+i\text{Z}\int_0^td\tau \frac{\Phi_0^{t-\tau}\left(\mathbf{R}_{ij};\mathbb{R}_{\overline{ij}};\tau\right)}{[4\pi i(t-\tau)]^{3/2}}+O(r_{ij})\label{BPMB},
\end{eqnarray}
where we have defined
\begin{eqnarray}
\text{Z}\int_0^td\tau \frac{\Phi_0^{t-\tau}\left(\mathbf{R};\mathbb{Z};\tau\right)}{[4\pi i(t-\tau)]^{3/2}}=\frac{1}{(4\pi i)^{3/2}}\lim_{\epsilon\rightarrow0^+}\left[\int_0^{t-\epsilon}d\tau \frac{\Phi_0^{t-\tau}\left(\mathbf{R};\mathbb{Z};\tau\right)}{(t-\tau)^{3/2}}-2\frac{\Phi(\mathbf{R};\mathbb{Z};t)}{\sqrt{\epsilon}}\right].
\end{eqnarray}
Finally, combining Eq. (\ref{Phi}), (\ref{psi}) and (\ref{BPMB}) we obtain an closed integral equation for $\Phi(\mathbf{R};\mathbb{Z};t)$
\begin{eqnarray}
-\frac{\Phi(\mathbf{R};\mathbb{Z};t)}{4\pi a(t)}=\psi_0(\mathbf{R},\mathbf{R},\mathbb{Z};t)+iZ\int_0^td\tau \frac{\Phi_0^{t-\tau}\left(\mathbf{R};\mathbb{Z};\tau\right)}{[4\pi i(t-\tau)]^{3/2}}+2\sum_{i=1}^{N-2}\psi_{\text{int}}(\mathbf{z}_i,\mathbf{R};\mathbb{\bar{Z}}_i;t)+\sum_{i<j}^{N-2}\psi_{\text{int}}(\mathbf{z}_i,\mathbf{z}_j;\mathbb{\bar{Z}}_{ij};t)\label{PhiIntegraleq}
\end{eqnarray}
where $\mathbb{\bar{Z}}_{i}=\mathbf{z}_1,\cdots,\mathbf{z}_{i-1},\mathbf{R},\mathbf{z}_{i+1},\cdots,\mathbf{z}_{N-2}$ and $\mathbb{\bar{Z}}_{ij}=\mathbf{z}_1,\cdots,\mathbf{z}_{i-1},\mathbf{R},\mathbf{z}_{i+1},\cdots,\mathbf{z}_{j-1},\mathbf{R},\mathbf{z}_{j+1},\cdots\mathbf{z}_{N-2}$.

In the case of spinless bosons starting from a pure non-interacting BEC state, we have $\psi(\mathbb{R},t=0)=1/V^{N/2}$ and thus $\psi_0(\mathbb{R},t)=1/V^{N/2}$. Below we will show that the ansatz $\Phi(\mathbf{R};\mathbb{Z};t)=\eta(t)/V^{N/2}$ provides an asymptotic exact solution in the region $\mathcal{D}_{\epsilon}$. Substituting this ansatz into Eq. (\ref{PhiIntegraleq}), we obtain
\begin{eqnarray}
-\frac{\eta(t)}{4\pi a(t)}-1-\hat{L}\eta(t)&=&2\sum_{i=1}^{N-2}\phi(\mathbf{z}_i-\mathbf{R},t)+\sum_{i<j}^{N-2}\phi(\mathbf{z}_i-\mathbf{z}_j,t),\label{phiansatz}\\
\phi(\mathbf{r},t)&=&i\int_0^td\tau G_0^{\text{rel}}(\mathbf{r},t\!-\!\tau)\eta(\tau).\label{phirt}
\end{eqnarray}
On a first sight, Eq. (\ref{phiansatz}) cannot be full filled because the l.h.s. only depends on $t$ but the r.h.s. is a function of $t,~\mathbb{Z}$ and $\mathbf{R}$. However, it is important to note that since $\int\phi(\mathbf{r},t)d^3\mathbf{r}=i\int_0^t\eta(\tau)d\tau$ is finite, $\phi(\mathbf{r},t)$ must decay very fast as $r\gg\sqrt{t}$. For example, for $a_\text{s}(t)=\beta\sqrt{t}$, we have
\begin{eqnarray}
\phi_{\beta}(\mathbf{r},t)&=&\frac{1}{1-\frac{2\sqrt{i}}{\sqrt{\pi}\beta}}\left[\text{Erf}\left(\frac{r}{2\sqrt{it}}\right)+\frac{2\sqrt{it}}{\sqrt{\pi}r}\exp\left(i\frac{r^2}{4t}\right) -1\right],
\end{eqnarray}
where $\text{Erf}(z)$  is the error function in the complex plane. In this case, we have $\phi_{\beta}(\mathbf{r},t)\rightarrow t^{3/2}/r^3$ as $r\gg\sqrt{t}$. For other form of $a_\text{s}(t)$, $\phi(\mathbf{r},t)$ decays even faster as $r\gg\sqrt{t}$. On the other hand, in region $\mathcal{D}_{\epsilon}$, we have $r_{ij}>\epsilon\gg\sqrt{t}$. This means the r.h.s. of Eq. (\ref{phiansatz}) vanishes in the region $\mathcal{D}_{\epsilon}$ and Eq. (\ref{phiansatz}) is reduced to Eq. (\ref{freeEta}) ( or Eq. (2) in the main text). As a result, the ansatz solution
\begin{eqnarray}
\Phi(\mathbf{R};\mathbb{Z};t)=\frac{\eta(t)}{V^{N/2}}\label{AnsatzPhi},
\end{eqnarray}
with $\eta(t)$ satisfying Eq. (\ref{freeEta}) becomes asymptotically exact in the region $\mathcal{D}_{\epsilon}$.

Finally, substituting Eq. (\ref{freeEta}) into (\ref{contact}) and the contact is given as
\begin{eqnarray}
\mathcal{C}(t)&=&\frac{1}{V}16\pi^2N(N-1)\int_{\mathcal{D}_{\epsilon}+\mathcal{I}_{\epsilon}} d^3\mathbf{R}d^3\mathbf{r}_3\,\cdots d^3\mathbf{r}_N\left|\frac{1}{4\pi}\Phi(\mathbf{R};\mathbf{r}_3\,\cdots\mathbf{r}_N;t)\right|^2\label{CtDplusI}\\
&=&\frac{1}{V}16\pi^2N(N-1)\int_{\mathcal{D}_{\epsilon}} d^3\mathbf{R}d^3\mathbf{r}_3\,\cdots d^3\mathbf{r}_N\left|\frac{1}{4\pi}\Phi(\mathbf{R};\mathbf{r}_3\,\cdots\mathbf{r}_N;t)\right|^2\label{CtD}\\
&=&\frac{1}{V}16\pi^2N(N-1)V^{N-1}\left|\frac{1}{4\pi}\frac{\eta(t)}{V^{N/2}}\right|^2\\
&=&n^2 |\eta(t)|^2,
\end{eqnarray}
where $V$ is the total volume of the system. The equality between Eq. (\ref{CtDplusI}) and (\ref{CtD}) is due to the fact that the contribution to the integral from the region $\mathcal{I}_{\epsilon}$ vanishes in the limit $\epsilon\ll 1/k_n$. The relative error caused by omitting the term $\int_{\mathcal{I}_{\epsilon}}\cdots$ is on the order of $k_n\epsilon$.

It is also straightforward to generalize above analysis to an initial state of arbitrary low energy plane wave state where
\begin{eqnarray}
\psi(\mathbf{r}_1,\cdots\mathbf{r}_N;t=0)&=&\frac{1}{N!V^{N/2}}\sum_{P}\exp\left(i\sum_{j=1}^N \mathbf{k}_{P_j}\cdot\mathbf{r}_j \right),\\
\psi_0(\mathbf{r}_1,\cdots\mathbf{r}_N;t)&=&e^{-iE_0t}\psi(\mathbf{r}_1,\cdots\mathbf{r}_N;0),\\
E_0&=&\frac{1}{2}\sum_{j=1}^N k_{j}^2,
\end{eqnarray}
where $P$ runs over all permutations. Here low energy means that $\max\{k_i\}\ll 1/\sqrt{t}$. In this case, one can show that
\begin{eqnarray}
\Phi(\mathbf{R};\mathbb{Z};t)=\eta(t)\psi_0(\mathbf{R},\mathbf{R},\mathbb{Z};t),
\end{eqnarray}
also provides an asymptotically exact solution in the region $\mathcal{D}_{\epsilon}$ and the contact is given by
\begin{eqnarray}
\mathcal{C}(t)&=&\frac{1}{V}N(N-1)\int_{\mathcal{D}_{\epsilon}} d^3\mathbf{R}d^3\mathbf{r}_3\,\cdots d^3\mathbf{r}_N\left|\psi(\mathbf{R},\mathbf{R},\mathbf{r}_3\,\cdots\mathbf{r}_N;0)\right|^2
|\eta(t)|^2\\
&=&\frac{1}{V}\int d^3\mathbf{R}\langle\Psi_0|\hat{\psi}^{\dag}(\mathbf{R})\hat{\psi}^{\dag}(\mathbf{R})\hat{\psi}(\mathbf{R})\hat{\psi}(\mathbf{R})|\Psi_0\rangle|\eta(t)|^2\\
&=&\frac{1}{V}\int d^3\mathbf{R}\langle\Psi_0|\hat{\psi}^{\dag}(\mathbf{0})\hat{\psi}^{\dag}(\mathbf{0})\hat{\psi}(\mathbf{0})\hat{\psi}(\mathbf{0})|\Psi_0\rangle|\eta(t)|^2\\
&=&g_2(0) |\eta(t)|^2,
\end{eqnarray}
which is Eq. (1) in the main text.

The above proof can be generalized straightforwardly to spin-1/2 fermions. In this case, we have
\begin{eqnarray}
\psi(\mathbf{r}_{1\uparrow},\cdots\mathbf{r}_{N\uparrow};\mathbf{r}_{1\downarrow},\cdots\mathbf{r}_{M\downarrow};t=0)&=&\frac{1}{N!M!V^{(N+M)/2}}\sum_{P_{\uparrow},P_{\downarrow}}(-1)^{P_{\uparrow}}(-1)^{P_{\downarrow}}\exp\left(i\sum_{j=1}^N \mathbf{k}_{P_{j\uparrow}}\cdot\mathbf{r}_{j\uparrow} + i\sum_{j=1}^M \mathbf{k}_{P_{j\downarrow}}\cdot\mathbf{r}_{j\downarrow} \right),\nonumber\\
\psi_0(\mathbf{r}_{1\uparrow},\cdots\mathbf{r}_{N\uparrow};\mathbf{r}_{1\downarrow},\cdots\mathbf{r}_{M\downarrow};t)&=&e^{-iE_0t}\psi(\mathbf{r}_{1\uparrow},\cdots\mathbf{r}_{N\uparrow};\mathbf{r}_{1\downarrow},\cdots\mathbf{r}_{M\downarrow};0),\\
E_0&=&\frac{1}{2}\sum_{j=1}^N k_{j\uparrow}^2+\frac{1}{2}\sum_{j=1}^M k_{j\downarrow}^2,
\end{eqnarray}
and following a similar derivation as given above for Bose gas, one can show that
the contact is given as
\begin{eqnarray}
\mathcal{C}(t)&=&\frac{1}{V}NM\int_{\mathcal{D}_{\epsilon}} d^3\mathbf{R}\prod_{\sigma=\uparrow,\downarrow}d^3\mathbf{r}_{2\sigma}\,\cdots d^3\mathbf{r}_{N\sigma}|\psi(\mathbf{R},\mathbf{r}_{2\uparrow},\cdots\mathbf{r}_{N\uparrow};\mathbf{R},\mathbf{r}_{2\downarrow},\cdots\mathbf{r}_{M\downarrow};0)|^2
|\eta(t)|^2\\
&=&\frac{1}{V}\int d^3\mathbf{R}\langle\Psi_0|\hat{\psi}^{\dag}_{\uparrow}(\mathbf{R})\hat{\psi}^{\dag}_{\downarrow}(\mathbf{R})\hat{\psi}_{\downarrow}(\mathbf{R})\hat{\psi}_{\uparrow}(\mathbf{R})|\Psi_0\rangle|\eta(t)|^2\\
&=&\frac{1}{V}\int d^3\mathbf{R}\langle\Psi_0|\hat{\psi}^{\dag}_{\uparrow}(\mathbf{0})\hat{\psi}_{\downarrow}^{\dag}(\mathbf{0})\hat{\psi}_{\downarrow}(\mathbf{0})\hat{\psi}_{\uparrow}(\mathbf{0})|\Psi_0\rangle|\eta(t)|^2\\
&=&g_2(0) |\eta(t)|^2,
\end{eqnarray}
which also gives Eq. (1) in the main text. Here for spin-1/2 fermions, $\mathcal{D}_{\epsilon}$ is defined as the region in which the distance between any $\uparrow$-particle and $\downarrow$-particle is larger than $\epsilon$.
\end{widetext}
\end{document}